# Electronic correlations in Fe at Earth's inner core conditions: effect of doping with Ni.

O. Yu. Vekilova[1], L. V. Pourovskii[2,3], I.A. Abrikosov[1,4] and S.I. Simak[1]

[1]Department of Physics, Chemistry and Biology, Linköping University, SE-58183 Linköping, Sweden
[2]Centre de Physique Théorique, CNRS, École Polytechnique, 91128, Palaiseau, France
[3] Swedish e-science Research Center, Department of Physics, Chemistry and Biology (IFM), Linköping University, Linköping, Sweden
[4]Materials Modeling and Development Laboratory, National University of Science and Technology 'MISIS', Moscow, Russia

## Abstract

We have studied the body-centered cubic (*bcc*), face-centered cubic (*fcc*) and hexagonal close-packed (*hcp*) phases of Fe alloyed with 25 at. % of Ni at Earth's core conditions using an *ab initio* local density approximation + dynamical mean-field theory (LDA+DMFT) approach. The alloys have been modeled by ordered crystal structures based on the bcc, fcc, and hcp unit cells with minimum possible cell size allowing for the proper composition. Our calculations demonstrate that the strength of electronic correlations on the Fe 3*d* shell is highly sensitive to the phase and local environment. In the bcc phase the 3*d* electrons at the Fe site with Fe only nearest neighbors remain rather strongly correlated even at extreme pressure-temperature conditions, with the local and uniform magnetic susceptibility exhibiting a Curie-Weiss-like temperature evolution and the quasi-particle lifetime $\Gamma$ featuring a non-Fermi-liquid temperature dependence. In contrast, for the corresponding Fe site in the hcp phase we predict a weakly-correlated Fermi-liquid state with a temperature-independent local susceptibility and a quadratic temperature dependence of $\Gamma$. The iron sites with nickel atoms in the local environment exhibit behavior in the range between those two extreme cases, with the strength of correlations gradually increasing along the hcp-fcc-bcc sequence. Further, the inter-site magnetic interactions in the bcc and hcp phases are also strongly affected by the presence of Ni nearest neighbors. The sensitivity to the local environment is related to modifications of the Fe partial density of states due to mixing with Ni 3d-states.

I. Introduction

Iron is the main component of Earth's inner core. From geochemical and seismic data it is generally assumed that iron in Earth's core is alloyed with 10-15 at. % of Ni. [1-2] The pressure and temperature inside the inner core are estimated to be ~6000 K and ~330-364 GPa, respectively. They are too extreme to be easily accessible in modern diamond anvil-cell experiments, so the actual crystal structure is still a matter of debate. From the discovery of Earth's core the hexagonal close-packed (*hcp*) structure has been the primary candidate. However, in the past decade a number of scientists started to advocate the body-centered cubic (*bcc*) Fe phase to be a suitable candidate for Earth's inner core material [3-4]. The *bcc* phase of Fe with 10 at. % of Ni has indeed been discovered experimentally at pressures and temperatures slightly lower than those of Earth's core [5], though this is still the only experimental indication of the existence of a high-temperature high-pressure *bcc* phase of Fe.

Despite a vast amount of publications, most of theoretical studies dedicated to the crystal structure and properties of Earth's core presumed Fe to be a non-magnetic metal with insignificant electronic correlations [3-4,6-7]. This picture can be justified by noticing that high pressure and high temperature, when considered separately, are both expected to suppress the correlations and, hence, to destroy the magnetism in Fe. Under high pressure the local Coulomb repulsion between the localized states ($U$) decreases due to the screening, while the bandwidth of these states ($W$) increases. The substantial reduction of the *U/W* ratio should lead to a decrease in correlations' strength thus justifying the use of standard mean-field approaches like the local density approximation (LDA) for the high-pressure studies. In addition, local magnetic moments are expected to be destroyed at high temperature by Stoner excitations.
Nevertheless, even the combined effect of extremely high pressure and temperature typical of Earth's core may not be sufficient to suppress significant electronic correlations in paramagnetic Fe. Pourovskii and co-workers have recently shown [8] that the pure bcc iron features a Curie-Weiss magnetic susceptibility and a non-Fermi-liquid behavior of the quasiparticle lifetime when local correlations are treated beyond the static mean-field level using dynamical mean-field theory (DMFT), while fcc and hcp Fe exhibit Fermi-liquid-like behavior. Ruban *et al.* [9] have predicted the existence of local magnetic moments of similar magnitude in all three (bcc, fcc, and hcp) phases of pure iron at Earth's core conditions using a longitudinal-fluctuation classical-spin model derived from disordered-local-moment density-functional-theory calculations.

However, as in particular has been noticed in Ref. [10], the presence of Ni in Earth's core, often neglected in first-principles studies, may be important for physical properties of the system. For example, the tendencies in relative stability of the bcc and hcp phases under pressure in pure iron and iron alloyed with nickel are opposite to each other [10]. Motivated by recent publications in this area we have studied in the present work the influence of Ni on electronic correlations in Fe-rich Fe-Ni alloys at the conditions of Earth's inner core. We have employed an *ab initio* LDA+DMFT approach to calculate the electronic structure of the bcc, fcc, and hcp phases for temperatures up to 5800 K and for densities corresponding to Earth's inner core. In the calculations we have used the volume equal to 7.05 Å for all the considered phases. This volume corresponds to the estimated density of Earth's inner core, 13155 kg/m$^3$, according to PREM [11].

We have made no attempt to match the experimentally reasonable Ni content (10-15 at. %), as well as the disordered state of Fe-Ni alloys expected at such extreme conditions [12]. They would require modeling with supercells, which are currently prohibitively large for *ab initio* DMFT calculations. Instead we have considered small $Fe_3Ni$ supercells (25 at. % of Ni) for all three phases. Despite their small sizes these ordered structures provided us with two types of the local environment for Fe atoms in the case of the bcc and hcp lattices, with and without Ni in the first coordination shell of Fe. Therefore, with this choice of supercells we have also been able to assess the impact of the local environment on electronic correlations between iron 3d states.

## II. Method

The calculations have been carried out using a self-consistent in the charge density approach [13-14] combining the full-potential linearized augmented plain-wave (FLAPW) Wien-2k [15] band structure method with a dynamical mean-field theory (DMFT) [16] treatment of the on-site Coulomb repulsion between the 3d states. The LDA+DMFT technique has previously been implemented to study the mechanical stability and magnetic properties of Fe [17-18].

For the on-site electron repulsion on Fe we have employed the parameters U=3.40 eV and J=0.90 eV in all the considered phases, following the evaluations in Ref. [8] based on the constrained random-phase approximation (cRPA) method [19-20], with the resulting Coulomb interaction matrices generated using the standard parameterization for the Slater integrals $U=F^0$, $J=(F^2+F^4)/14$, and $F^2/F^4=0.625$ [21]. In Ref. 8 the U value has been found to be quite similar in the different phases of Fe, varying from 3.04 eV for the fcc to the 3.37 eV for the hcp crystal structures. Our test calculations of the dependence of the magnetic susceptibility and quasiparticle lifetime on the change of the U for the $Fe_3Ni$ system showed that the small differences in U (about 10 %) did not induce any qualitative change of our results (see Figs. 1-2, and Figs. A1-A3 in Appendix). Moreover, those small differences may stem from a disentanglement procedure employed within cRPA, which is an approximation. U on Fe may also be affected by doping with Ni. Due to all those uncertainties we believe it is reasonable to neglect small differences in the value of U between the phases. Hence, we fixed U to the largest value obtained for the pure Fe phases, i.e. we took 3.4 eV for all the considered phases. For Ni in $Fe_3Ni$ we chose the same value as for Fe, which can be considered as an upper bound for the magnitude of U there. The calculated value of U in pure Ni is smaller than in Fe [22], however, nickel sites in $Fe_3Ni$ have a very different environment compared to that of pure Ni. Hence, no additional effects of correlations can be expected for Ni beyond those discussed below.

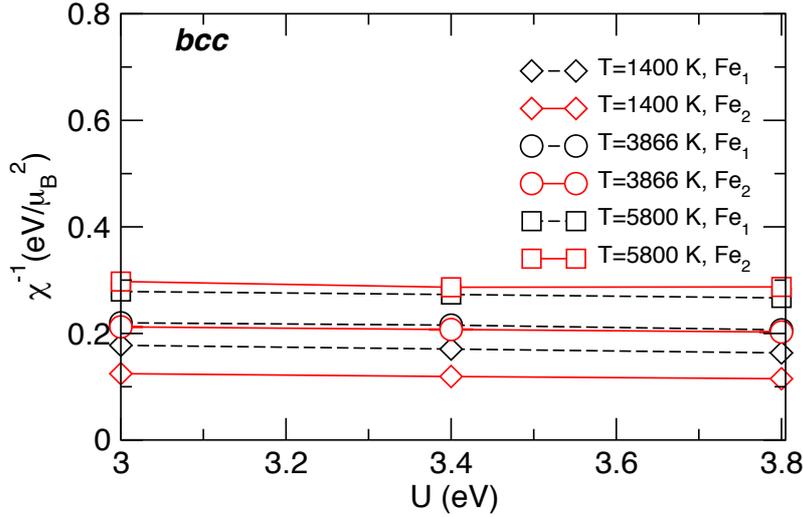

Fig. 1 (Color online) The dependence of the inverse uniform magnetic susceptibility on the values of U, varying from 3 to 3.8 eV for different temperatures in the bcc phase of $Fe_3Ni$. The solid red lines correspond to the $Fe_2$ type of Fe atoms; the dashed black ones correspond to the $Fe_1$ type. Diamonds, circles and squares are used for the temperatures 1400 K, 3866 K and 5800 K, respectively. As one may see, nearly horizontal lines mean that the increasing or decreasing of U by 10% does not quantitatively affect the results, independently on the local environment around the Fe atoms (for both $Fe_1$ and $Fe_2$ types of atoms; the details on the atomic arrangement are described in the Sec. II).

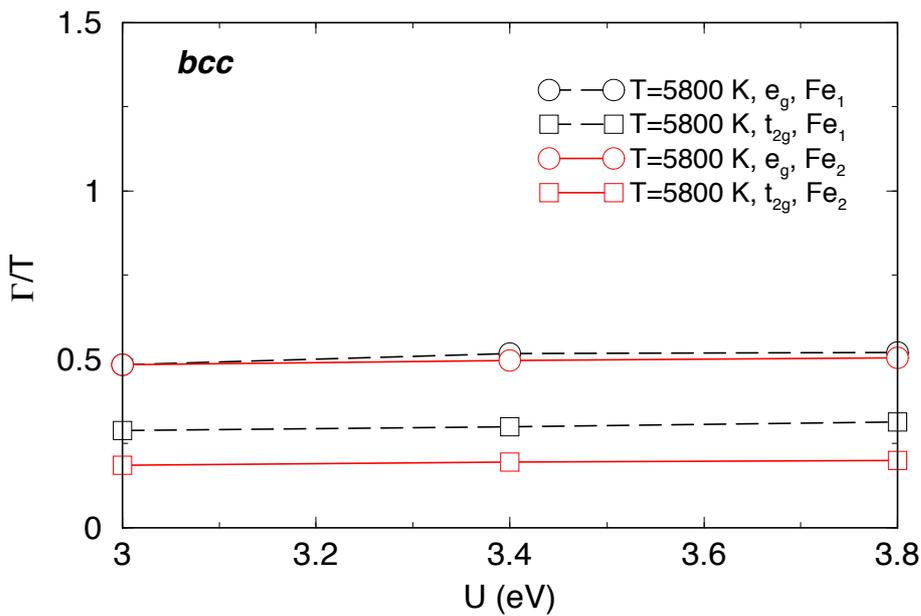

Fig. 2. (Color online) The dependence of the quasiparticle lifetime on the value of U for the temperature 5800 K in the Fe$_3$Ni bcc phase. The solid red lines correspond to the Fe$_2$ type of Fe atoms; the dashed black ones correspond to the Fe$_1$ type. Circles and squares are used for the e$_g$ and t$_{2g}$, respectively. As one may see, increasing or decreasing the value of U by 10% does not quantitatively change the results, independently of the local environment around the Fe atoms.

More details of the calculations can be found in Ref. [8]. The DMFT quantum impurity problem has been solved using the Quantum Monte-Carlo (CT-QMC) method [23] as implemented in the TRIQS package [24], using 5 x 10$^8$ CT-QMC moves with the measurements performed after each 200 moves. For the Coulomb repulsion parameter we have used the density-density Slater parameterization.

The uniform magnetic susceptibility has been computed as the reaction of the system to a small magnetic field applied on all sites. This susceptibility might be calculated for each site from the resulting on-site magnetic moment. The uniform magnetic susceptibility consists of the response to the field at a given site (i.e. its local susceptibility) as well as of the interaction with the moments induced by the same field on other sites. An external magnetic field of H = 0.005 eV/µ$_B$ along the $\hat{z}$ axis has been added to the one-electron Hamiltonian in order to compute the uniform magnetic susceptibility $\chi = \frac{M}{H}$ from the resulting small magnetic moment *M*.

The static local susceptibility is a reaction of the system to a magnetic field applied to a given site only. This local susceptibility has been calculated via the spin-spin correlation function (computed during the CT-QMC calculations), to which it is related by the fluctuation-dissipation theorem [25]. The local magnetic susceptibility $\chi_{loc} = \int_0^{1/T} \chi(\tau)\, d\tau$ has been computed from the imaginary-time on-site spin-spin correlation function $\chi(\tau) = \langle \hat{S}_Z(\tau)\hat{S}_Z(0)\rangle$ sampled by the CT-QMC method. Hence, by comparing the local and uniform susceptibilities of a given site one may extract the magnitude of its exchange interaction with its neighbors.

In order to calculate the quasiparticle lifetime, the imaginary part of self-energy $\Sigma$ for the first several frequencies has been fitted with a polynomial and extrapolated to the zero frequency (see Fig. 3). The imaginary-frequency self-energy has been analytically continued to the real axis using a stochastic maximum-entropy approach [26]. The derivative has been taken in order to calculate $Z_{eff}$ (the effective mass) at zero frequency.

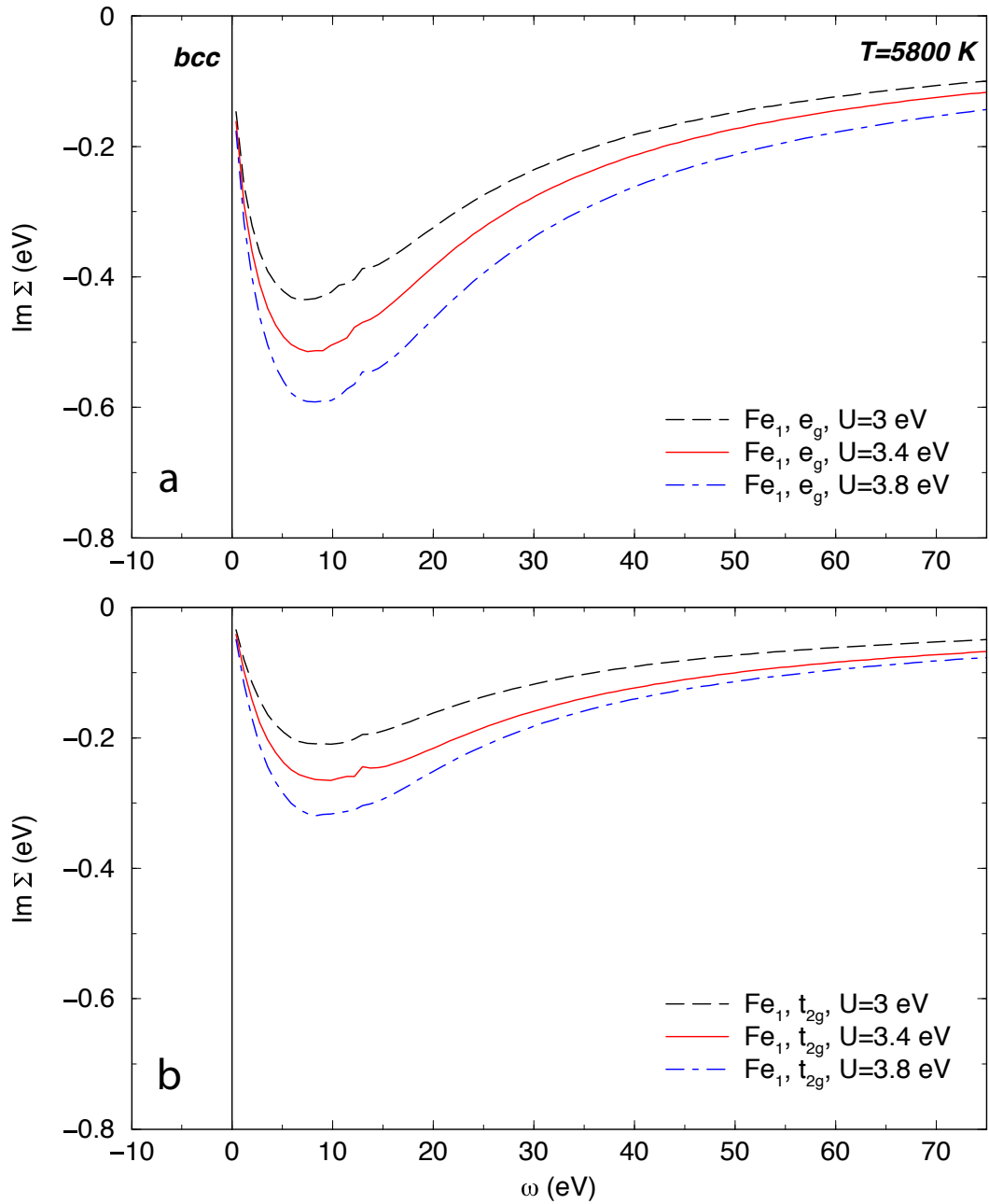

Fig. 3. (Color online) The dependence of the imaginary part of self-energy on the frequency for different values of U for the bcc phase of $Fe_3Ni$: $e_g$ (a) and $t_{2g}$ (b) states. Temperature is equal to 5800 K. Black dashed, red solid and blue dashed-dotted lines correspond to the U values of 3 eV, 3.4 eV and 3.8 eV respectively. As one may see, all three curves are close to each other at low-frequencies, meaning that the corresponding values for the mass

enhancement and inverse quasi-particle life-time exhibit a rather weak dependence on variations of U within this range.

We used the standard double counting for metals, i.e. weakly correlated systems, the around mean-field expression [27]. As shown in Ref. [14] the LDA+DMFT electronic structure exhibits rather weak dependence on the choice of the double-counting when the self-consistency in the charge density is included. 1x1x2 bcc and hcp-based cubic and hexagonal unit cells with 3 Fe atoms and 1 Ni atom have been used to simulate bcc and hcp Fe-Ni alloys, respectively. The c/a ratio for the hcp phase has been equal to 1.6 [28]. $L1_2$ structure with 3 Fe and 1 Ni atom has been used to simulate fcc Fe-Ni alloy. Accordingly, there are two non-equivalent Fe atoms in the case of the bcc and hcp phases, due to the differences in the local environment. $Fe_1$ type corresponds to the iron atom whose nearest neighbors are 4 (6) iron and 4 (6) nickel atoms in the bcc (hcp) structure, respectively. The second iron atom of $Fe_2$ type is surrounded by exclusively iron atoms in the first coordination shell. For the fcc cell all the three Fe atoms are equivalent, so one deals only with the case where the local environment of Fe atoms contains Ni.

## III. Results.

The temperature evolution of the LDA+DMFT inverse magnetic susceptibility has been calculated for temperatures in the range from 1000 K to 5800 K and is shown in Fig. 4. In the case of the bcc phase (Fig. 4b) the inverse magnetic susceptibility of the $Fe_2$ type (red solid line) exhibits strong and linear temperature dependence. This behavior can be described by the Curie-Weiss law and hints at the presence of a local magnetic moment on this site. This result is in good agreement with Ref. [8], where a Curie-Weiss temperature evolution has been obtained for pure bcc iron. Hence, the $Fe_2$ atoms that have only Fe nearest neighbors exhibit magnetic properties similar to those of pure bcc iron. In contrast, the inverse magnetic susceptibility of the $Fe_1$ atoms (black dashed line) shows a weaker and non-linear temperature dependence, which can be expected for a Pauli paramagnet at high temperature. This difference points out to a significant reduction in electronic correlations on Fe sites that have Ni nearest neighbors.

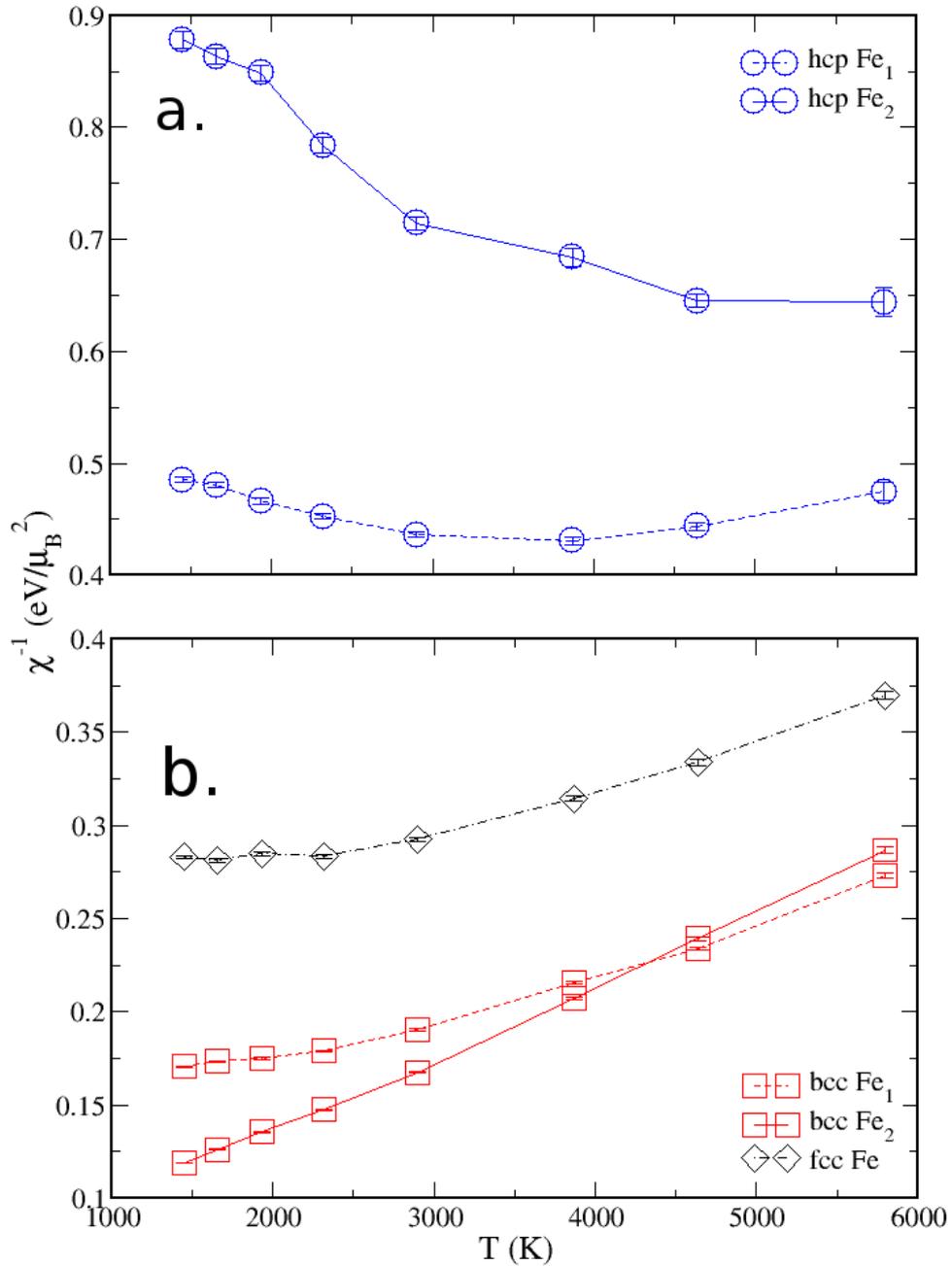

Fig 4. (Color online) (a). The inverse uniform magnetic susceptibility in paramagnetic state as a function of temperature for the hcp phase, with the dashed and solid lines corresponding to the $Fe_1$ (six Ni and six Fe nearest neighbors) and the $Fe_2$ (all nearest neighbors are Fe) types, respectively. The error bars show the stochastic CT-QMC error.

(b). The same data for the bcc and fcc phases of $Fe_3Ni$ alloy. Dashed red line corresponds to the $Fe_1$ type of iron atoms, whose nearest neighbors are 4 Fe and 4 Ni atoms in the bcc structure. The solid red line shows the $Fe_2$ type of atoms, which are surrounded exclusively by Fe atoms. The dashed-dotted black line shows the Fe atoms of the fcc phase. The error bars show the stochastic CT-QMC error.

The magnetic susceptibility of Ni itself is small and exhibits rather weak temperature dependence (See Fig. 5). Similar LDA+DMFT calculations have been carried out for the hcp and fcc phases of Fe$_3$Ni (Fig. 4a and b, respectively), and non-linear and rather weak temperature dependence of the magnetic susceptibility has been obtained. The magnetic susceptibility of Ni sites (Fig. 5) in those phases exhibits a similar temperature evolution. Total uniform magnetic susceptibility for all the considered phases is presented in Fig. 6.

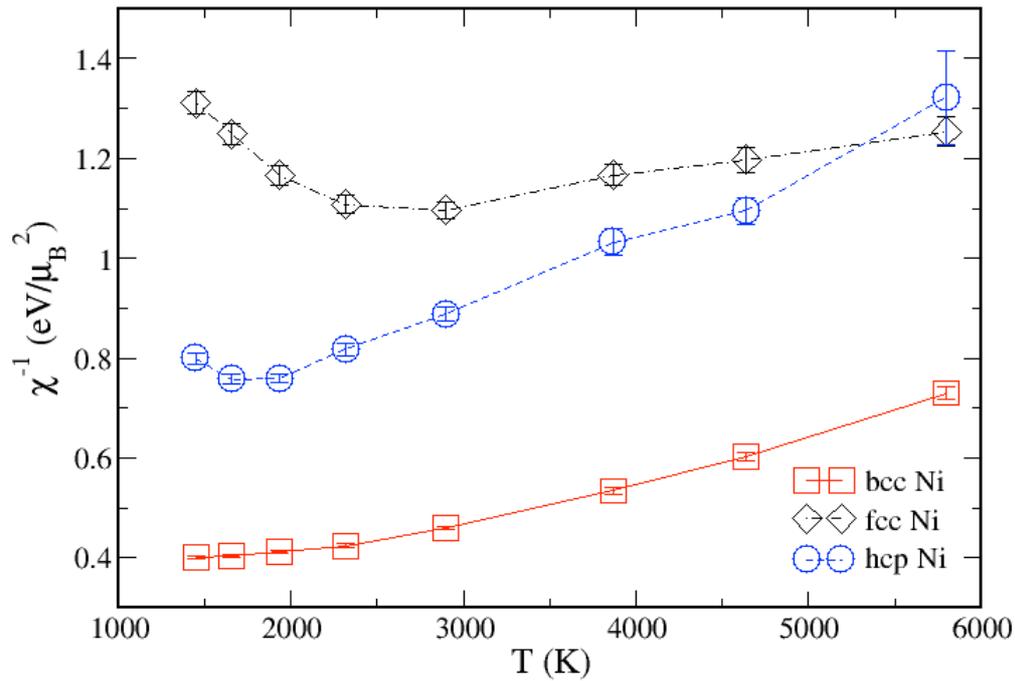

Fig 5. (Color online) The inverse uniform magnetic susceptibility of the paramagnetic state as a function of temperature for the Ni atom in the bcc, fcc and hcp phases of Fe$_3$Ni alloy. The solid red line corresponds to the Ni atoms in bcc structure of Fe$_3$Ni alloy, the dashed-dotted black line shows the Ni atoms in the fcc structure and the blue dashed line corresponds to the Ni atoms in the hcp structure. The error bars show the stochastic CT-QMC error.

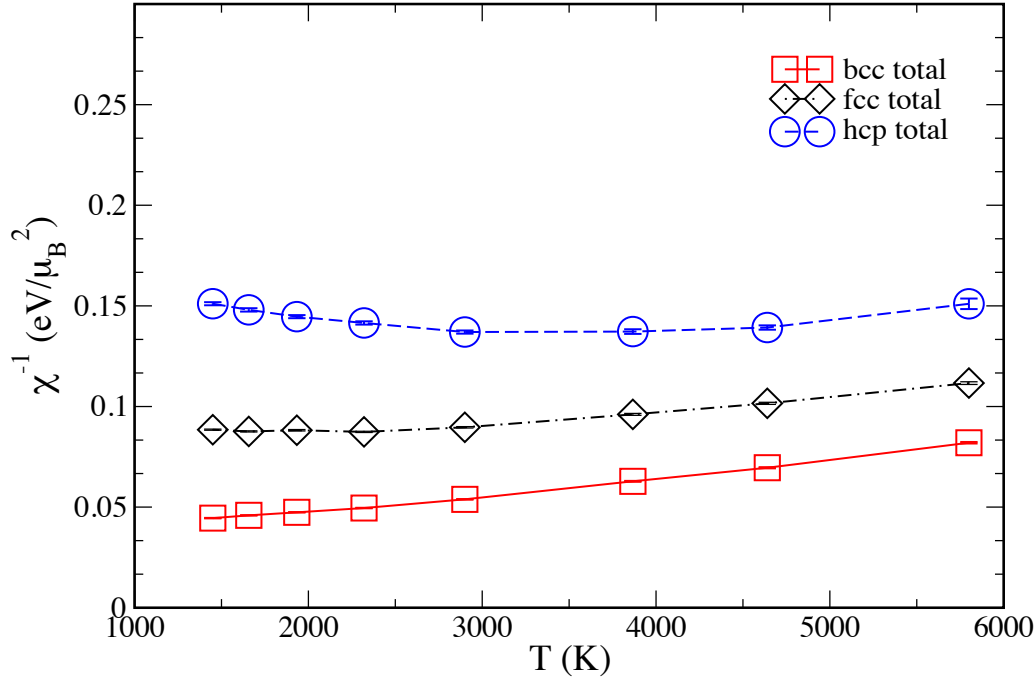

Fig. 6. (Color online) The inverse total uniform magnetic susceptibility of the paramagnetic state as a function of temperature for the Ni atom in the bcc, fcc and hcp phases of $Fe_3Ni$ alloy. The solid red, dashed-dotted black and dashed blue lines correspond to the total magnetic susceptibility of the bcc, fcc and hcp structures of $Fe_3Ni$ alloy, respectively. The error bars show the stochastic CT-QMC error.

The inverse of magnetic susceptibility of the $Fe_2$ atoms in the bcc phase has been fitted to the Curie-Weiss law:

$$\frac{1}{\chi} = \frac{3(T+\Theta)}{\mu_{eff}^2} \tag{1}$$

The calculated effective local magnetic moment, $\mu_{eff}$, at the volume corresponding to the pressure ~ 300 GPa is $2.6\mu_B$ and $\Theta$ = 1347 K. The parameter $\Theta$ here comprises two terms: a characteristic coherence temperature, around which a crossover from the correlated high-temperature state to a low-temperature Fermi-liquid one takes place [29] and a contribution from inter-site magnetic interactions (see, e. g., Ref. [16]).

In order to extract the contribution of inter-site interactions to the uniform susceptibility we have also calculated the imaginary-time spin-spin correlation functions and evaluated the corresponding static local magnetic susceptibilities (i.e. response functions to a static on-site magnetic field) $\chi_{loc}$. In a mean-field theory the local and uniform susceptibilities for a given site are related as

$$\chi^{-1} = \chi_{loc}^{-1} - J_0, \tag{2}$$

where $J_0$ is the sum of all inter-site exchange interactions of this site with the other sites. Calculated inverse $\chi_{loc}$ of Fe sites vs. temperature is shown in Fig. 7a. One may notice that the local susceptibility of Fe is very sensitive to the phase and local environment. It is virtually temperature-independent for the $Fe_2$ type of the hcp phase, the behavior characteristic for a weakly-correlated itinerant Fermi liquid. A linear temperature evolution of $1/\chi_{loc}$ for the $Fe_2$ site in bcc $Fe_3Ni$ is expected for correlated systems with local moments. $\chi_{loc}$ of other Fe types are in between those two extremes: inverse local susceptibilities of the $Fe_1$ site in bcc and Fe in fcc feature clear deviations from linearity, while $\chi_{loc}$ of $Fe_1$ site in hcp exhibits a moderate decrease with temperature.

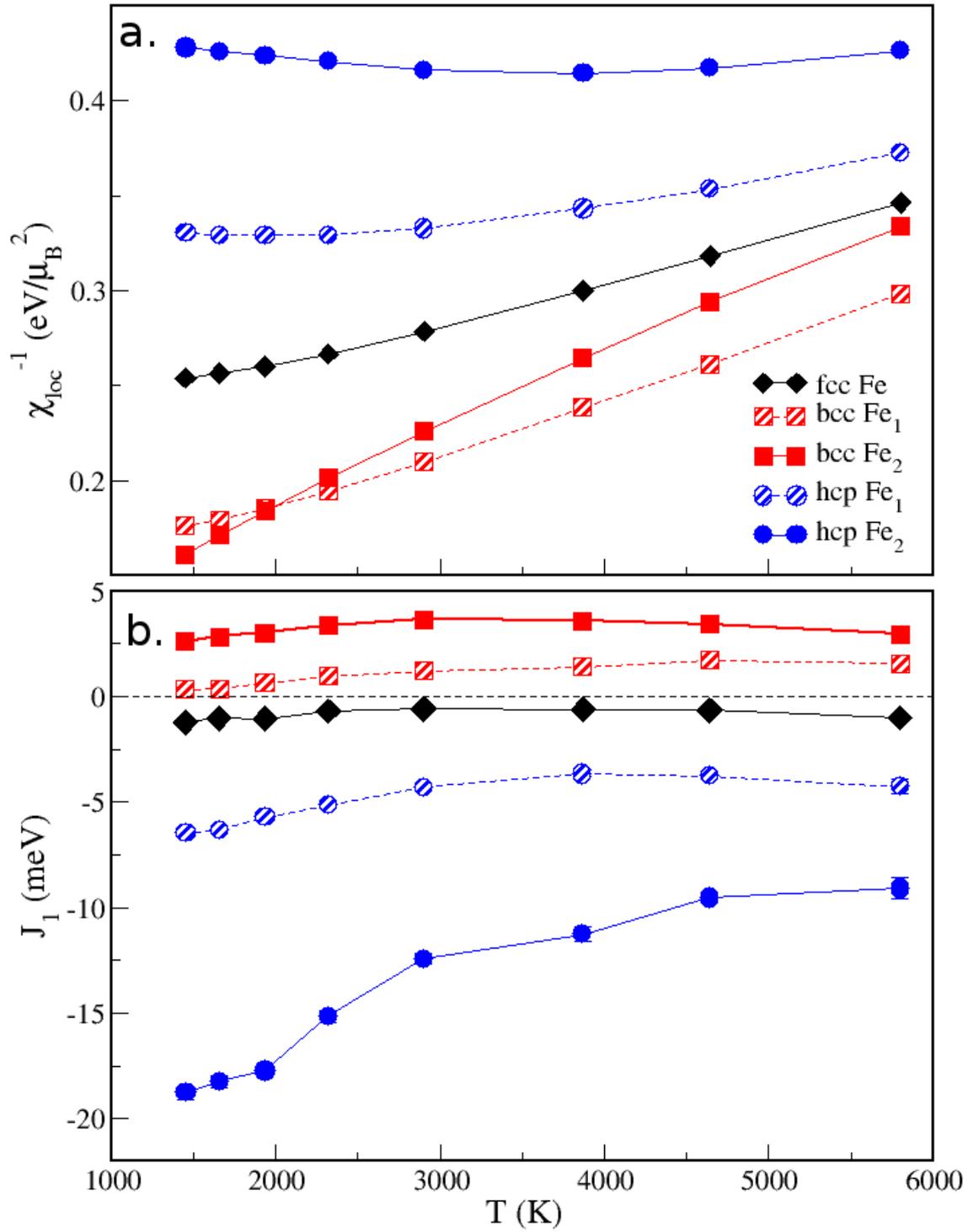

Fig. 7 (Color online). (a). The inverse local susceptibility of Fe atoms in the bcc, fcc, and hcp $Fe_3Ni$ phases vs. temperature. The notation for iron sites ($Fe_1$, $Fe_2$) is the same as in Fig. 1.

(b). The average inter-site nearest-neighbor exchange interaction as function of temperature. The positive/negative values correspond to ferromagnetic/antiferromagnetic interactions, respectively.

From this analysis we conclude that the qualitative difference in temperature evolution of the uniform susceptibilities is mainly due to on-site correlations captured by $\chi_{loc}$, in contrast to the results of Ref. [9]. However, the contribution of the inter-site interactions to the actual magnitude of uniform susceptibility is also significant, as one may notice by comparing corresponding $\chi$ and $\chi_{loc}$. We have evaluated $J_0$ using eq. (2) and then have extracted the average nearest-neighbors inter-site interaction $J_1$ assuming that longer-range interactions are negligible (this assumption is not very reasonable for the weakly-correlated itinerant fcc and hcp phase, however, we plot the inter-site interactions in this form to allow for a more simple comparison with the data on $J_1$ obtained by other methods). Then $J_0 = 2\sum_i J_{0i} \sim 2zJ_1$, where $i$ runs over the nearest neighbors of a given site 0, $z$ is the coordination number equal to 12 and 8 for fcc/hcp and bcc, respectively.

The resulting interaction $J_1$ vs. temperature is plotted in Fig. 7b. $J_1$ favors ferromagnetic ordering in the bcc phase and anti-ferromagnetic one in the fcc and hcp phases, in agreement with the well-known tendencies of the corresponding pure Fe phases at lower pressures-temperatures. One may expect that a large and rather strongly temperature-dependent $J_1$ for Fe$_2$ in the hcp phase in fact comprises significant contributions from longer-range anti-ferromagnetic interactions making this system highly frustrated. One may also notice that the presence of Ni nearest neighbors leads to drastic reduction of inter-site correlations in both the bcc and hcp phases.

We have also evaluated the strength of correlations in all three phases by analyzing the low-frequency behavior of the DMFT self-energy. The degree of non-Fermi-liquid behavior can be estimated from the inverse quasiparticle lifetime, which reads:

$$\Gamma = -Z\Im[\Sigma(i0^+)], \qquad (3)$$

where the quasiparticle residue $Z$ is calculated as:

$$Z^{-1} = 1 - \frac{\partial \Im\Sigma(i\omega)}{\partial \omega}\Big|_{\omega \to 0^+}, \qquad (4)$$

from the imaginary self-energy, $\Im\Sigma$, extrapolated to zero. $\omega$ stands for the frequency.

The temperature evolution of the inverse quasiparticle lifetime $\Gamma/T$ for the relevant irreducible representations of the Fe 3d shell of all three phases is shown in Fig. 8.

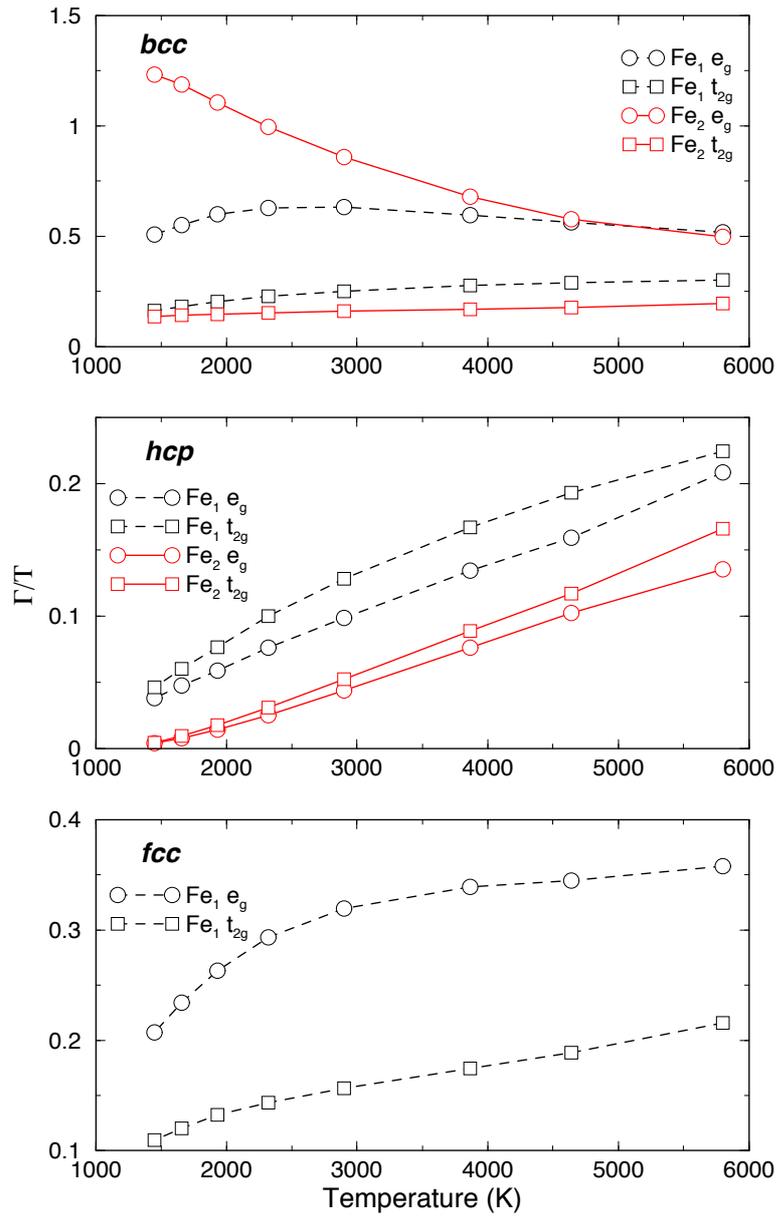

Fig 8. (Color online) The ratio of the inverse quasiparticle lifetime Γ to temperature T versus T for the bcc, fcc and hcp phases of the Fe$_3$Ni alloy. The dashed black lines correspond to e$_g$ (circles) and t$_{2g}$ (squares) of the Fe$_1$ type. The solid red lines correspond to e$_g$ and t$_{2g}$ of the Fe$_2$ type, respectively.

In the Fermi-liquid regime the inverse quasiparticle lifetime scales as $T^2$ and, therefore, a linear dependence of $\Gamma/T$ vs. temperature is expected. In the bcc phase the $e_g$ states of $Fe_1$ (dashed black line) and $Fe_2$ (solid red line) behave in a quite different way. At high temperatures (higher than 1000 K) $\Gamma/T$ of the $Fe_2$ atoms exhibits substantial deviations from linearity and decreases with temperature, indicating the non-coherent nature of these states. We note that the presence of a local magnetic moment at the $Fe_2$ site inferred from the Curie-Weiss temperature dependence of its magnetic susceptibility is consistent with this manifestly non-Fermi-liquid behavior of the corresponding quasi-particle lifetime. As discussed in Ref. [8], the Curie-Weiss-like behavior of magnetic susceptibility can also be explained within a Fermi-liquid picture as being due to the large peak at the Fermi energy in the DOS of the bcc phase. However, such an explanation cannot simultaneously account for the non-Fermi-liquid temperature dependence of the quasi-particle lifetime. $\Gamma/T$ of the $Fe_1$ atoms for the $e_g$ states increases with temperature up to around 2400 K and the deviations from the linear regime begin at higher temperatures. This behavior is intermediate between the strongly non-linear $e_g$ states of $Fe_1$ and the quasi-linear dependence observed for weakly-correlated states, e. g., for bcc $t_{2g}$. The $t_{2g}$ states of both $Fe_1$ and $Fe_2$ in the bcc phase, as well as $e_g$ states of the fcc Fe show rather small deviations from the linear regime and thus from the Fermi-liquid behavior. In the hcp phase the behavior of $\Gamma/T$ versus temperature is almost perfectly linear for all Fe sites and representations, as expected for a Fermi liquid.

In Figs. 9 and 10 we display the partial densities of states (PDOS) of Fe and Ni in all three $Fe_3Ni$ phases obtained within LDA+DMFT and LDA, respectively.

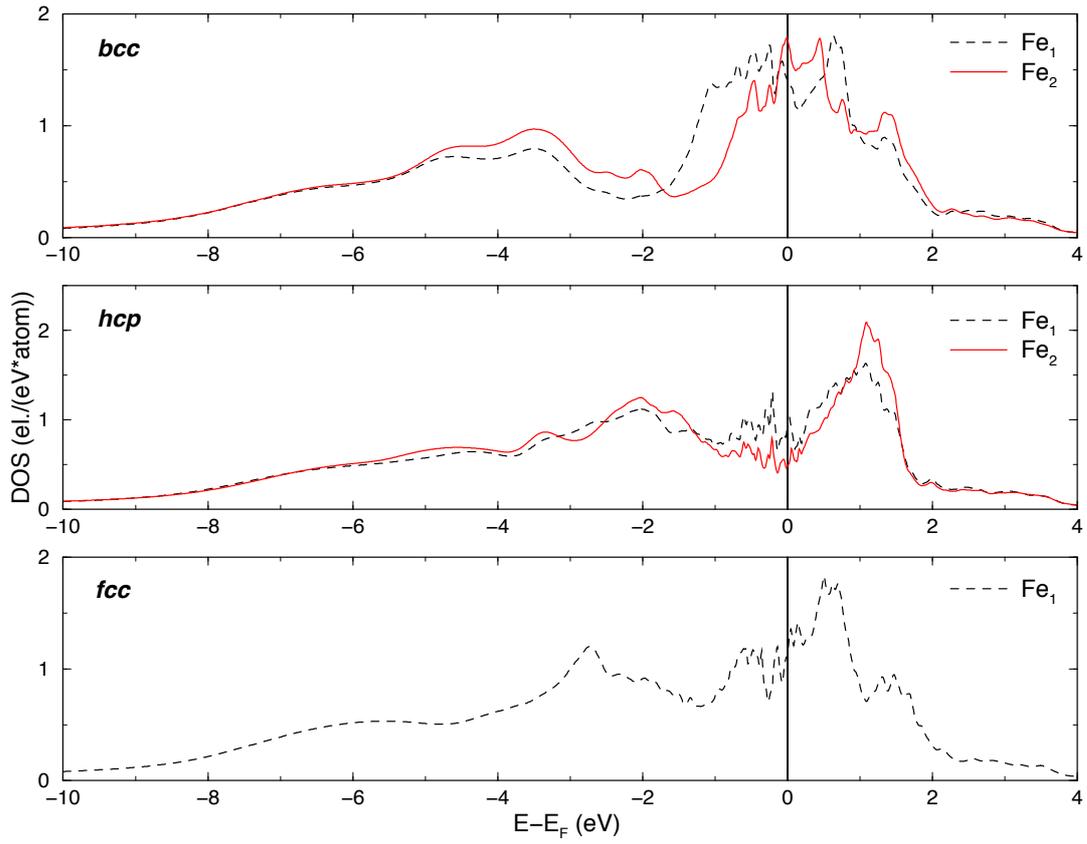

Fig 9. (Color online) The LDA+DMFT partial DOS of iron for the bcc, fcc and hcp phases of $Fe_3Ni$ alloy. The dashed black and solid red lines are the partial DOS of the $Fe_1$ type and $Fe_2$ types, respectively.

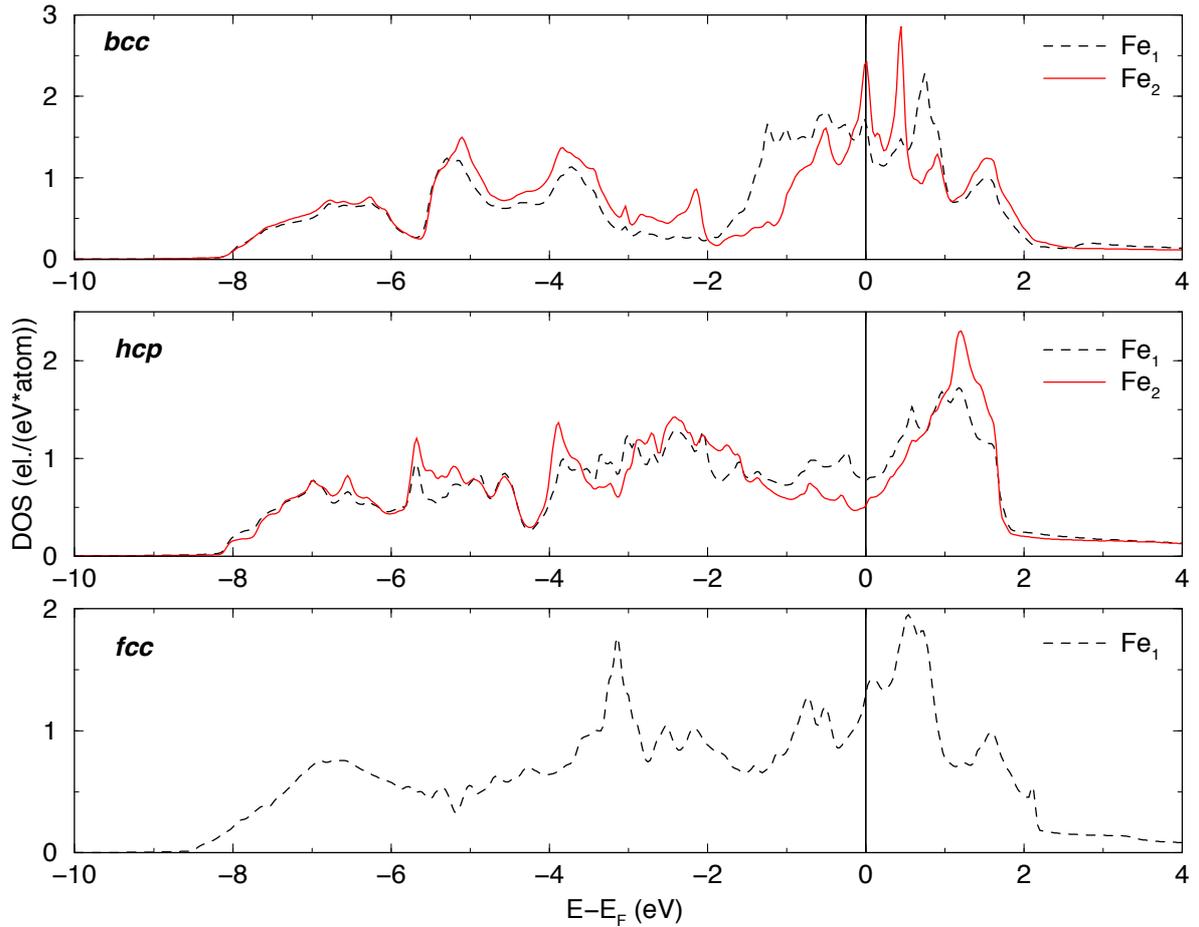

Fig 10. (Color online) The LDA partial DOS of iron for the bcc, fcc and hcp phases of Fe$_3$Ni alloy. The dashed black and solid red lines are the partial DOS of the Fe$_1$ type and Fe$_2$ types, respectively.

By comparing the corresponding LDA and LDA+DMFT PDOS one may observe a noticeable quasi-particle renormalization in bcc and fcc, while the hcp PDOS is almost unaffected by correlations. A prominent peak is present at the Fermi level $E_F$ of the LDA Fe PDOS of bcc Fe$_3$Ni. The width of this peak at half-maximum differs in the Fe$_1$ and Fe$_2$ cases, with the Fe$_2$ peak being significantly narrower. A narrow peak at $E_F$ is associated with rather strong correlations in bcc Fe as has been previously discussed in Refs.[8,30-31]. Thus, the d-electrons of Fe$_1$ become less correlated due to the effect of Ni nearest neighbors, which induce a broadening of the peak at $E_F$, while the overall d-band bandwidths remains similar in both Fe$_1$ and Fe$_2$.

## IV. Conclusions

We have carried out a theoretical study of the role of electronic correlations in the Fe$_3$Ni alloy at pressures ~ 300 GPa and temperatures up to 5800 K corresponding to Earth's inner core conditions within an *ab initio* LDA+DMFT approach.

Our calculations have demonstrated strong sensitivity of electronic correlations and the iron magnetism in Fe$_3$Ni to the phase and local environment. In the bcc phase the Fe sites with exclusively Fe nearest neighbors feature a Curie-Weiss uniform susceptibility corresponding to a local magnetic moment of 2.6 $\mu_B$, while the Fe sites with 4 Ni nearest neighbors exhibit a weakly temperature-dependent Pauli-like susceptibility. The temperature evolution of the local susceptibility is also strongly affected by the phase and environment. It ranges from a strong inverse-temperature dependence observed in the bcc phase for the Fe sites with Fe-only nearest neighbors to an almost constant Pauli-like behavior for the corresponding Fe sites in hcp. The inter-site exchange interactions are ferromagnetic in bcc and antiferromagnetic in hcp and fcc. In the bcc and hcp they are strongly affected by the presence of Ni nearest neighbors.     A similar sensitivity to the phase and local environment has been observed for the quasi-particle lifetime in the bcc phase, with the overall picture being in perfect agreement with the one inferred from the analysis of the magnetic susceptibility. The sensitivity to the local environment is explained by broadening of the Fe partial DOS at the Fermi level due to mixing with itinerant states of its Ni neighbors, which remain weakly correlated. Correspondingly, in real iron-rich Fe-Ni bcc alloys, if they are stable at the inner Earth core conditions, one may expect strong variations in the correlations' strength between different Fe sites due to variations in the local environment. As in the case of pure iron [8], the Fe$_3$Ni fcc and hcp phases behave as rather weakly-correlated Fermi liquids, however, the inter-site antiferromagnetic correlations in hcp are also significantly affected by the local environment.

**Acknowledgements**

We are thankful to HPC-Europa2 for support of this project. LVP acknowledges travel support provided by PHD DALEN 2012 under the project 26228RM. Support from the Swedish Research Council (VR) Projects No. 621-2011-4426, 2011-42-59; LiLi-NFM, the Swedish Foundation for Strategic Research (SSF) program SRL10-0026, the Swedish Government Strategic Research Area Grant in Materials Science "Advanced Functional Materials" (AFM) and Swedish e-Science Research Centre (SeRC) are gratefully acknowledged. IAA acknowledges support by Grant of Russian Federation Ministry for Science and Education (grant No. 14.Y26.31.0005). The computations were partly performed on resources provided by the Swedish National Infrastructure for Computing (SNIC).

# Appendix

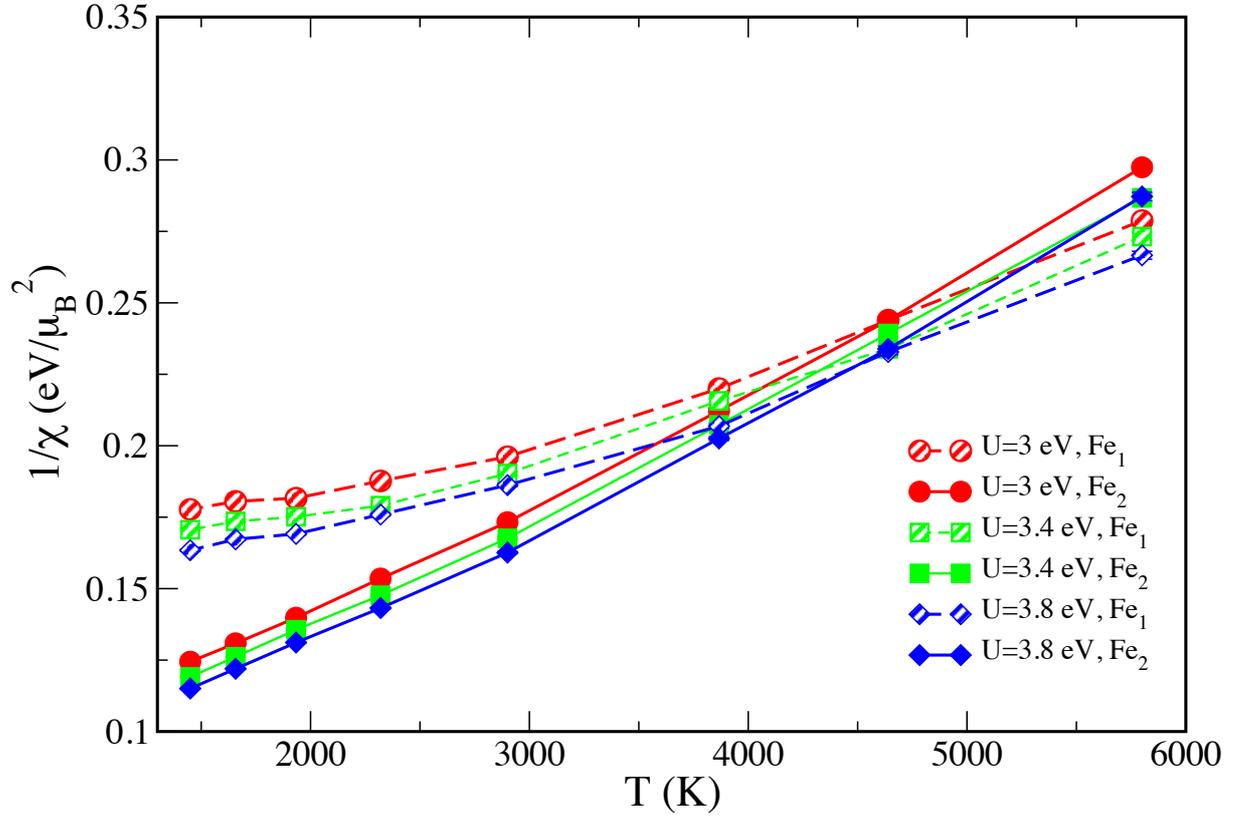

Fig. A1 (Color online) The dependence of the inverse uniform magnetic susceptibility on temperature for values of U, varying from 3 to 3.8 eV in the bcc phase of $Fe_3Ni$. The solid and dashed lines correspond to the $Fe_2$ and $Fe_1$ type of Fe atoms, respectively. Red circles, green squares and blue diamonds are used for the U values 3, 3.4 and 3.8 eV, respectively. As one may see, the dependence of the susceptibility on temperature during the increasing or decreasing of U by 10% does not quantitatively affect the results, independently on the local environment around the Fe atoms.

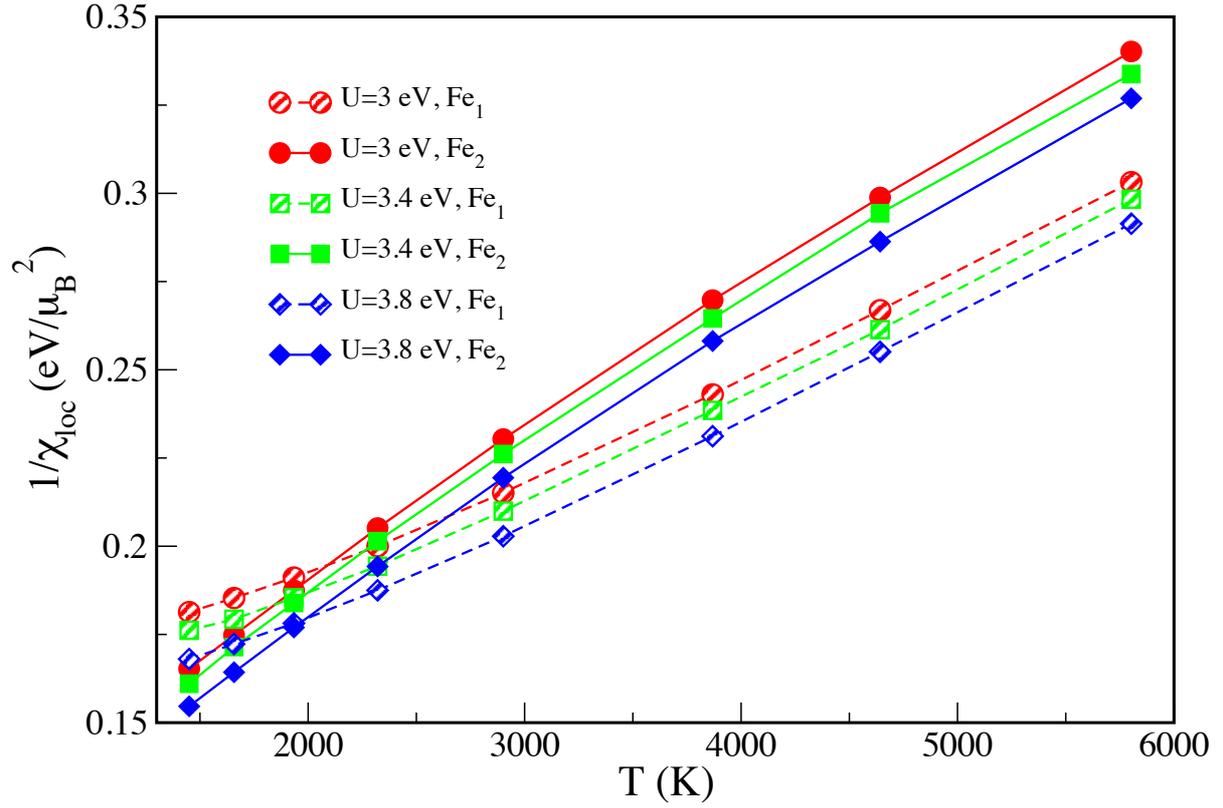

Fig. A2 (Color online) The dependence of the inverse local magnetic susceptibility on temperature for values of U, varying from 3 to 3.8 eV in the bcc phase of $Fe_3Ni$. The solid and dashed lines correspond to the $Fe_2$ and $Fe_1$ type of Fe atoms, respectively. Red circles, green squares and blue diamonds are used for the U values 3, 3.4 and 3.8 eV, respectively. As one may see, the dependence of the susceptibility on temperature during the increasing or decreasing of U by 10% does not quantitatively affect the results, independently on the local environment around the Fe atoms.

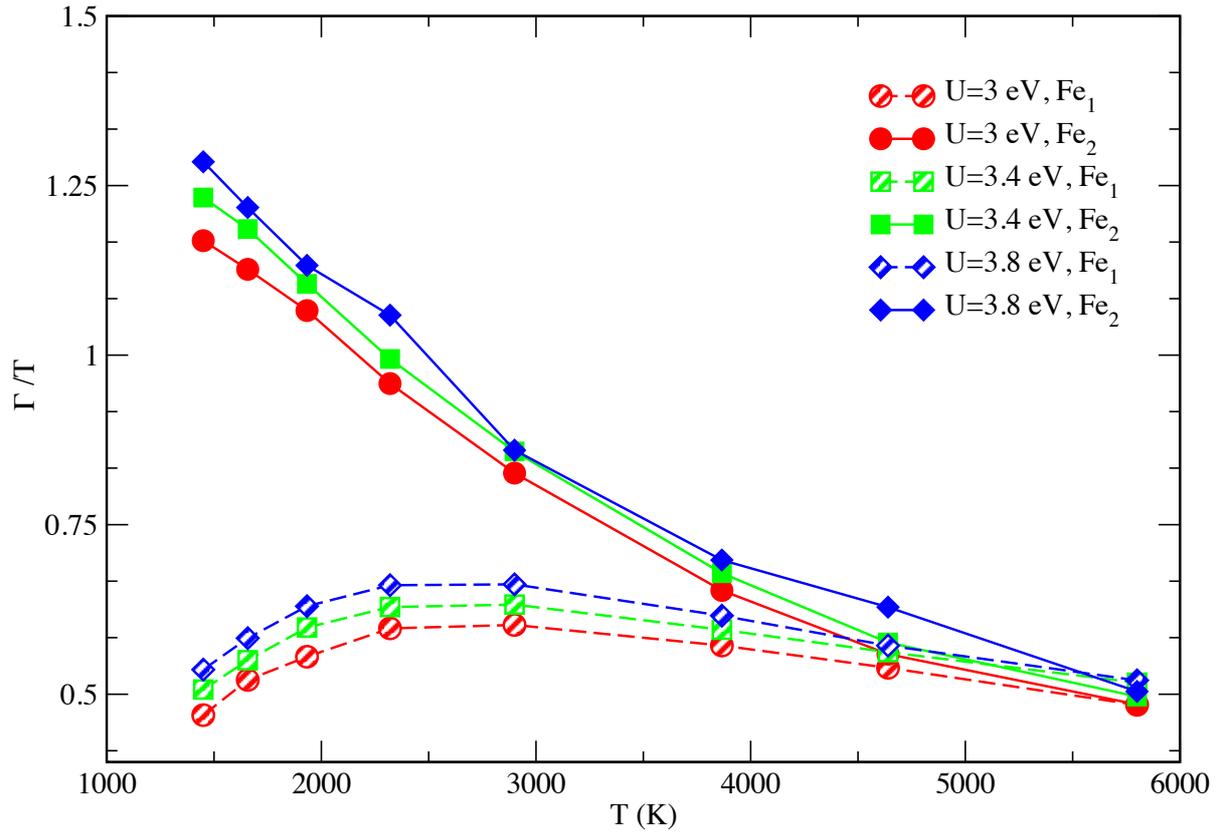

Fig. A3 (Color online) The dependence of the inverse quasiparticle lifetime Γ to temperature T versus T for values of U, varying from 3 to 3.8 eV in the bcc phase of $Fe_3Ni$. The solid and dashed lines correspond to the $Fe_2$ and $Fe_1$ type of Fe atoms, respectively. Red circles, green squares and blue diamonds are used for the U values 3, 3.4 and 3.8 eV, respectively. As one may see, the dependence of Γ/T on temperature during the increasing or decreasing of U by 10% does not quantitatively affect the results, independently on the local environment around the Fe atoms.